\documentclass[iop]{emulateapj}

\usepackage{epstopdf}
\usepackage{graphics}
\usepackage{natbib}

\newcommand{\sgra}{\mbox{$\rm{Sgr A}^{*}$}}

\newcommand{\msun}{\mbox{$M_{\odot}$}}

\newcommand{\kms}{\mbox{km s$^{-1}$}}

\newcommand{\etal}[1]{{ et al.}~}
\def\kms{\ifmmode \hbox{km~s}^{-1}\else km~s$^{-1}$\fi}
\def\etal {{\it et al.}}
\def\deg      {{\ifmmode^\circ\else$^\circ$\fi} } 

\def\h2     {H$_2$}

\def\arcsec{\hbox{$^{\prime\prime}$}}

\shorttitle{G2}
\shortauthors{Scoville \& Burkert}

\date{~~~~~~~~~~~~~~~~~~~~~~~~~~~~~~~~~~~~~~~~~~~Accepted ApJ 2/25/13}                                           

\begin{document}

\title{The Galactic Center Cloud G2 -- a Young Low-Mass Star with a Stellar Wind}
 \author{ N. Scoville\altaffilmark{1}, 
and  A. Burkert\altaffilmark{2,3,4}}

\altaffiltext{1}{California Institute of Technology, MC 249-17, 1200 East California Boulevard, Pasadena, CA 91125}
\altaffiltext{2}{University Observatory Munich, Scheinerstrasse 1, D-81679 Munich, Germany}
\altaffiltext{3}{Max-Planck-Fellow}
\altaffiltext{4}{Max-Planck-Institute for Extraterrestrial Physics, Giessenbachstrasse 1, 85758 Garching, Germany}


\altaffiltext{}{}

\begin{abstract}
We explore the possibility that the G2 gas cloud falling in towards \sgra is the mass loss envelope of a young TTauri star. 
As the star plunges to smaller radius at 1000 to 6000 \kms, a strong bow shock forms where the stellar wind is impacted by the 
hot X-ray emitting gas in the vicinity of \sgra. For a stellar mass loss rate of $4\times10^{-8}$ \msun~ per yr and wind velocity 100 \kms, 
the bow shock will have an emission measure ($EM = n^2 vol$) at a distance $\sim10^{16}$ cm, similar to that inferred from the 
IR emission lines.  The ionization of the dense bow shock gas is potentially provided by collisional ionization at the shock front 
and cooling radiation (X-ray and UV) from the post shock gas. The former would predict a constant line flux as a function of 
distance from \sgra, while the latter will have increasing emission at lesser distances. In this model, the star and its mass loss wind should survive pericenter passage since 
the wind is likely launched at $0.2$ AU and this is much less than the Roche radius at pericenter ($\sim3$ AU for a stellar mass of 2\msun). In this model, the emission cloud 
will probably survive pericenter passage, discriminating this scenario from others.

\end{abstract}

\section{Introduction}

The recently discovered G2 cloud which is infalling toward  \sgra is a most intriguing 
astronomical discovery \citep{gil12a} -- both its origin and nature are unclear as yet. Nevertheless,  in the space of a few years from the first 
detection, one will observe its passage within $\sim2200$ Schwarzschild  radii of the supermassive 
black hole -- in September 2013 \citep{gil13}. Numerous observations, from radio to X-ray, are planned 
for this 'once in an astronomical lifetime' event. At this point it is unclear if the cloud will survive pericenter 
passage and whether the activity of \sgra will increase and over what timescale.

G2 was first observed in the HI Br$\gamma$ and Br$\delta$ HI and 2.058$\mu$m HeI emission lines
and detected in the near infrared continuum with an extremely low 550K color temperature \citep{gil12a,eck13}. Over the period 2004 to 2012 
its 3d velocity has increased from 1200 to over 2500 \kms \citep{gil13}. The latest orbital determination indicates an 
eccentricity of 0.966 and pericenter passage at $2\times10^{15}$ cm from \sgra, when the 3d velocity will be 6340 \kms \citep{gil13}. The orbital period 
is 198 yrs with an apocenter distance of $1.6\times10^{17}$ cm and velocity 108 \kms.  

The observed flux in the
Br$\gamma$ line requires an ionized gas emission measure $EM = \int n_e^2~dvol \sim10^{57}$ cm$^{-3}$. Surprisingly, the 
line flux exhibits no change greater than 10\% over the 4 yr period \citep{gil12a,gil13}. The cloud is resolved along
its orbital path but unresolved in the transverse direction (i.e. $\leq 10^{15}$ cm); \cite{gil13} adopt an effective spherical radius $1.88\times10^{15}$ cm for the emitting region and thereby deduce a mean density of $6\times10^5$ cm$^{-3}$ and a total mass $\sim3$ M$_{earth}$ (assuming unity 
volume filling for the ionized gas). If this is the whole story (i.e. G2 has only the mass seen in the ionized gas and it is uniformly 
distributed), then it is clear that the cloud can not survive pericenter passage since the Roche limit for tidal stability is $n \sim 1.5\times10^{17}$ cm$^{-3}$.

Several models have been proposed for the origin and nature of G2. \cite{gil12a}, \cite{sch12} and \cite{bur13} have suggested that it may 
have formed as an interstellar cloud from colliding stellar winds in the young stellar ring at $2\times10^{17}$ cm radius.  \cite{mey12} suggest it is a ring of gas 
formed by a Nova explosion. These explanations might  
account for the low angular momentum and high eccentricity orbit. \cite{bur13} model the subsequent hydrodynamic evolution of the cloud as it falls toward \sgra -- yielding reasonable agreement with the observed emissivities and kinematic evolution. Alternatively, \cite{mur11} 
proposed that G2 is a star with a protoplanetary disk, also scattered out of the young stellar ring (presumably from a triplet star system). Then 
as the system descends towards \sgra, the disk is photo-evaporated and tidally disrupted to produce
the G2 cloud. With the orbital parameters known at the time they proposed this model, they argued that the outer protoplanetary disk at 5 - 10 AU would probably survive 
pericenter passage. However, with the most recent orbit determination \citep{gil13}, the disk will now probably be  tidally stripped to within 2 AU radius at pericenter. In both of the above scenarios, it seems we are then 
extraordinarily fortunate to be observing a one-off event (i.e. a single orbital event) only noticed within a few years 
of its final demise. And yet there do not appear to be large numbers of similar objects further out. 

Here we explore a different scenario -- G2 being a young low mass TTauri star, formed in the young stellar ring and subsequently injected into the 
eccentric orbit. Many TTauri stars have mass-loss winds at 200 - 500 \kms with $\dot{M} \sim1 - 5 \times 10^{-8}$\msun~yr$^{-1}$ during their first million years. The mass-loss 
rates for those with measured rates ($\sim50$\% of the sample) have a very large range $10^{-6.5 ~\rm{to}~ -10}$\msun~yr$^{-1}$ \citep{har95} and a median value $2.5\times10^{-9}$\msun~yr$^{-1}$. 
\cite{whi04} obtained median values $10^{-7.0}$ and $10^{-8.2}$\msun~yr$^{-1}$ for samples of 8 and 42 class I and II TTauri stars, respectively. 
These outflows may originate as a centrifugally driven wind from the inner accretion disk \citep[e.g.][]{bla82}. If so, then the observed velocities imply a launch radius well inside 1 AU radius from the star. 
This scenario for G2 has a superficial similarity to that of  \cite{mur11} in having a young stellar object formed 
in the young stellar ring and having a circumstellar disk, but is very different in the physics of the mass-loss material 
and the possibility of pericenter survival. In the case of the TTauri star wind, the outflow velocities $\sim100$\kms are much larger than the 10\kms expected for a
photo-evaporating disk with velocities $\sim10$\kms. For the TTauri star wind, a  very dense bow shock is formed at radius $\sim10^{14}$cm and it is this gas which produces the
observed emission lines.  The stellar wind plus bow shock model readily reproduce the observed emission line 
fluxes using standard TTauri star wind parameters. 

In the following we analyze the mass-loss wind parameters and the interaction with the hot X-ray emitting gas in the vicinity 
of \sgra, followed by a detailed numerical model for the bow shock at the upstream side of the plunging star. This allows us to 
track the evolution of the bow shock structures and their emissivity as a function of distance from \sgra. Throughout most of the orbit, tidal stripping is not 
very significant since the bow shock on the front side of the star 
is pushed to smaller radii from the star as the 
orbit approaches the central black hole (due to the higher stellar velocity and higher density of the ambient hot X-ray gas at small galactic radii). 

Using the derived density structure, 
we then analyze the ionization of the gas, concluding 
that the most likely source of the observed emission lines is the dense bow shock where the outflowing stellar wind meets the 
$10^{8-9}$K shocked layer of ambient gas. The ionization may be provided by photons in the free-free continuum and line emission of the gas cooling behind the shock. 
In addition, there will be collisional ionization as the wind material passes through the shock at 200 - 500 \kms. The expected Lyman continuum from young 
stars in the central parsec does not appear to be adequate. If the ionization is collisional it would account for the apparent 
constancy of the emission line fluxes, since the mass-loss rate is probably constant; however, unless the mass-loss rate is 
$> 10^{-7}$ \msun~yr$^{-1}$ this ionization is insufficient to account for the inferred emission measures. 

\section{Spherical Mass-Loss Wind and Interaction with Coronal X-ray Emitting Gas}

As the low mass star with a stellar wind descends towards \sgra, the outer envelope will interact with the ambient hot X-ray emitting gas \citep{bag03,mun04,shc10}. 
For the stellar wind, we adopt the following fiducial numbers :
\begin{equation}
\dot{M} = 4\times10^{-8}\msun \rm{yr}^{-1} \equiv \dot{M}_{*}~ 
\end{equation}\label{mdot} 
and
\begin{equation}
V_{W} = 100~ \rm{\kms}.
\end{equation}

\noindent which implies a density distribution in the mass-envelope,  

\begin{equation}
\rho_W = { \dot{M} \over{4\pi r^2 V_W}} = 9.6 \times10^{-17}  { \dot{M}_{*} \over{r_{AU}^2 {V_W}  _{100}}} \rm{gr~cm^{-3}} 
\end{equation}\label{rho_w}
\begin{equation}
n_W  \simeq 6\times 10^7 \rm{~H~cm^{-3}} \rm{~~at ~1 ~AU. }
\end{equation}

\noindent For the stellar wind we adopt an inner launch radius which scales with the wind velocity, assuming the outflow velocity is equal to
the escape velocity for material initially in circular orbit at the launch radius. This is a good approximation for outflow where the radial acceleration 
is gradual (rather than explosive).

\begin{figure}[ht]
\epsscale{1.}
\plotone{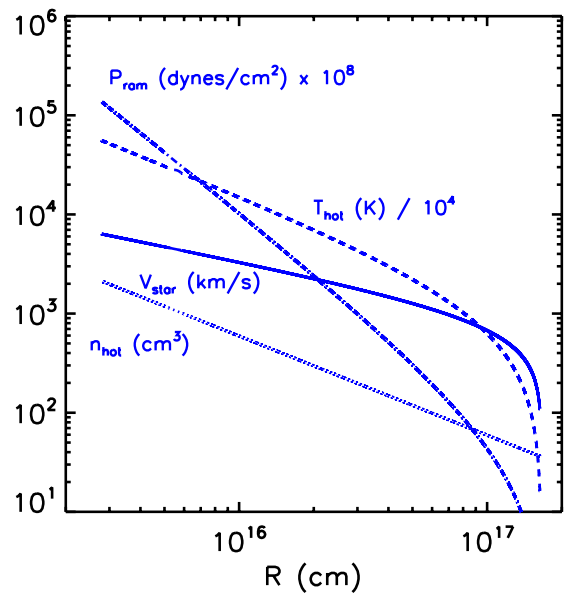}
\caption{The stellar 3d velocity as a function of distance from \sgra.  The adopted density, temperature and ram pressure 
of the hot X-ray emitting gas are also shown. }\label{vstar}
\end{figure}

For the hot ambient medium, we use the radial density and temperature distributions given by \cite{bur13}, which are similar but not identical to  the X-ray 
model of \cite{yua03}. The temperature distribution was assumed appropriate to hydrostatic equilibrium of the gas in the potential of the
$4.3\times10^6$ \msun~ \sgra ~black hole.  Specifically, we adopt :

\begin{equation}
\rho_{hot} = 9.5\times10^{-22} \left( { {10^{16}\rm{cm}} \over{r}} \right) ~ \rm{gr~cm^{-3}} 
\end{equation}\label{hot_X-ray}
and
\begin{equation}
T_{hot} = 2\times10^{8}  \left( { {10^{16}\rm{cm}} \over{r}}\right)~ \rm{K}.
\end{equation}
\noindent The above ignores the possibility that some of the X-ray emission is from stellar sources \citep{saz12}. These 
distributions are shown in Fig \ref{vstar} together with the 3d stellar velocity and the resultant ram pressure from the hot X-ray emitting gas.

\begin{figure}[ht]
\epsscale{1.}
\plotone{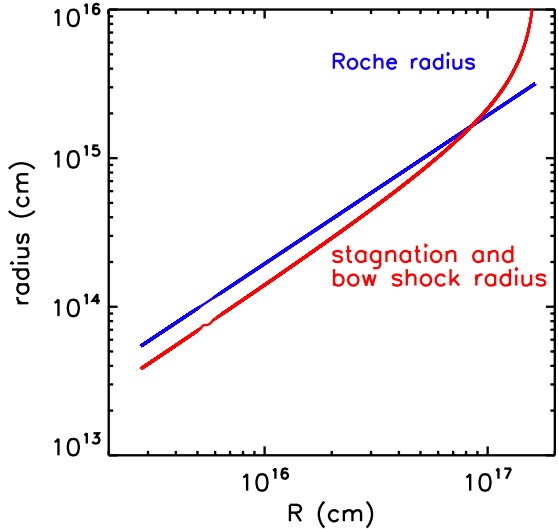}
\caption{The stagnation radius in front of the star (Eq. \ref{rstag}) and the Roche radius for tidal stripping are shown as a function of distance from \sgra.  The Roche radius was calculated assuming a star mass of 2\msun.   Inside $10^{17}$ cm radius from \sgra, the emission region is stable against tidal disruption, although tides will shear the tail out behind the star.}\label{radius}
\end{figure}

The outflowing stellar wind will terminate on the upstream side of the star in a shock front at the point where its ram pressure equals the thermal pressure of the hot ambient
gas or the ram pressure of the hot gas. In the portion of the orbit where G2 is currently observed, it is plunging with velocity 
significantly greater than the local circular, virial velocity; thus the ram pressure is likely to dominate the hot gas thermal pressure (since the thermal temperature 
of this gas was estimated assuming virial equilibrium in the central potential, which is dominated by the black hole at these radii). 

\noindent  The bow shock of the stellar wind against the hot medium will have a stagnation or standoff radius in front of the star 
(balancing ram pressures) at :

\begin{equation}
R_s^2 = { \dot{M}_* V_W \over{ 4\pi \rho_H V_*^2} }
\end{equation}
where $\rho_H \sim 10^{-21}$ gr cm$^{-3}$ is the mass density in the hot medium at an orbital distance $\sim10^{16}$cm, corresponding to the
position of G2 in mid 2012. For mid 2012, $V_*$ = 2000 \kms and therefore 

\begin{equation}
R_s = 2.3 \times 10^{14}~\rm{cm}  \left[ { \dot{M}_{*}  {V_W}  _{100} \over {{\rho_H}_{-21} {V_*}_{2000}}} \right]^{1/2} \sim 14~\rm{AU}.\label{rstag}
\end{equation}

\noindent At this radius the wind density is ${n_W}_0= 3 \times 10^5$ H cm$^{-3}$.

For the mass-loss star moving supersonically through the hot medium near \sgra, the bow shock on the 
upstream side will occur at $R_s$  and a conical compressed gas layer will extend downstream from the star. In fact there will be two shocked layers, 
the first where the ambient hot gas meets the stellar wind bow shock, and the second, an interior bow shock, where the outflowing stellar wind meets
the compressed gas at the stagnation point on the upstream side of the star. We will refer to these as the {\bf hot bow shock} and the {\bf cold bow shock} respectively (Fig. \ref{model}). 
The immediate post shock gas temperatures
are given by $T = 1.38\times10^5 (V_{shock} / 100 ~\kms)^2$~K for an ionized gas \citep{mck80}, implying temperatures of $10^{8-9}$ and $\sim5\times10^5$~K, respectively,   
behind the two bow shocks.  

The first shock front will be adiabatic 
since this very hot gas cools slowly. The density in this shock is shown in Fig. \ref{density} and is $\sim1000$ cm$^{-3}$. The second shock front has a very high post shock density ($\sim10^{8}$cm$^{-3}$), allowing it to cool in just $10^{-3}$ yrs (comparable to the sound crossing time for the cold bow shock). This shock will be modeled very approximately as isothermal. Interior to this second, isothermal shock which is a very thin sheet of gas, lies the free streaming stellar wind.

In the context of this model, there are clearly several locations from which observed ionized emission lines might arise: 1) the high 
density, interior, cold bow shock; 2) the stellar mass-loss envelope below the bow shock and 3) the outer, hot bow shock. The latter is not a significant source 
 given the very high temperatures (and hence low recombination rate) and its relatively small emission measure ($n^2 L$). To evaluate the expected emissivities from 
the mass-loss envelope and the cold bow shock, we develop a detailed model 
for the bow shock structure in \S\ref{bow_shock_model} and the ionization in \S\ref{ionization}. 

\section{Bow Shock Model}\label{bow_shock_model}

\begin{figure}[h]
\epsscale{1.}
\plotone{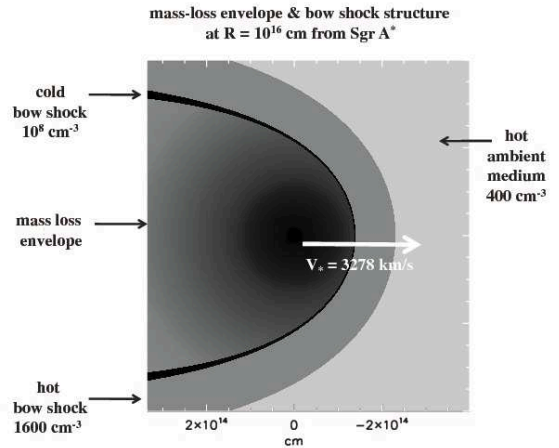}
\caption{The computed density structure of the mass-loss envelope and bow shocks are shown when the star is $10^{16}$cm from Sgr A$^*$. The density 
is shown with logarithmic scaling from n = 100 to 10$^{10}$ cm$^{-3}$ in grayscale. The stellar velocity at this radius is 3278 km/s and the ambient medium density
of the X-ray emitting gas is 400 cm$^{-3}$. Two bow shocks are formed -- a thick one in the shocked hot upstream medium and a very thin, high density
cold layer in the shocked stellar wind. For this model, $V_{wind} = 100$ \kms and $\dot{M} = 4\times10^{-8}$ \msun~  yr$^{-1}$}\label{model}
\end{figure}

To model the bow shock region, we make use of the analytic treatment by \cite{dys75} modeling a fast stellar wind ablating an expanding dust globule. 
Dyson developed two limiting cases : 1) with mixing of the two shocked layers (hot gas and cold gas) and 2) without mixing of the shocked layers,
resulting in a full tangential discontinuity between the layers. In the following we use the latter approximation. The 3d velocity of the star relative to the ambient
medium is taken from the latest orbit of G2 given by \cite{gil13} and the ambient medium density and temperature as a function of 
distance from \sgra were taken from Eq. \ref{hot_X-ray} and all are shown in Fig. \ref{vstar}. 

Inside $\sim5\times10^{16}$cm distance from \sgra, the mass-loss star is plunging supersonically toward \sgra. At these radii the star 
is essentially in free-fall, whereas the hot gas is assumed to be in hydrostatic equilibrium (hence having thermal sound speed similar to the circular velocity). 
The supersonic motion of the star through the hot gas will result in the two bow shocks mentioned earlier: the cold bow shock in the stellar wind material just inside the stagnation radius (see Fig \ref{radius}), 
and the hot bow shock in the hot ambient medium just outside the stagnation radius. Figure \ref{model} shows these shocks and the stellar envelope structure as computed for 
a distance of $10^{16}$ cm from \sgra when the star is moving at 3300 \kms through the ambient medium of density $\sim440$ cm$^{-3}$. The cold bow shock has a thickness of only $\sim3\times10^{12}$
at this point and so it is hardly visible in Fig. \ref{model} but its density is 10$^8$ cm$^{-3}$, so its potential emission measure ($n^2 L$) is large. We say 'potential' since it is also required that the
gas be ionized if it is to account for the observed line emission. 

\begin{figure}[ht]
\epsscale{1.}
\plotone{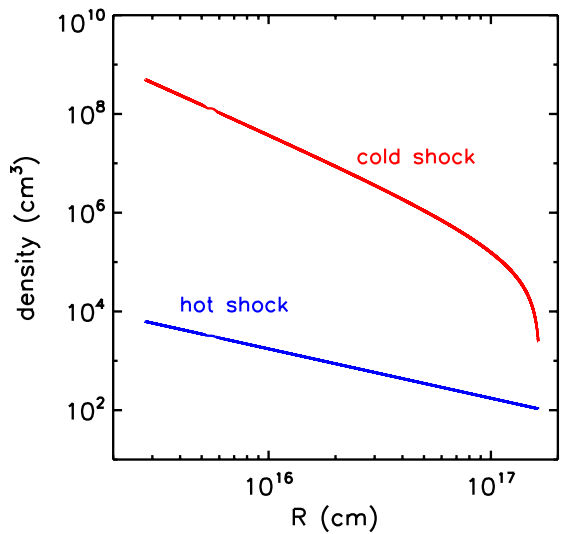}
\caption{The density of the cold and hot bow shocks are shown as a function of distance from \sgra.
[The density 
in the cold shock is independent of $V_W$ and $\dot{M}$.]}\label{density}
\end{figure}

\begin{figure}[ht]
\epsscale{1.}
\plotone{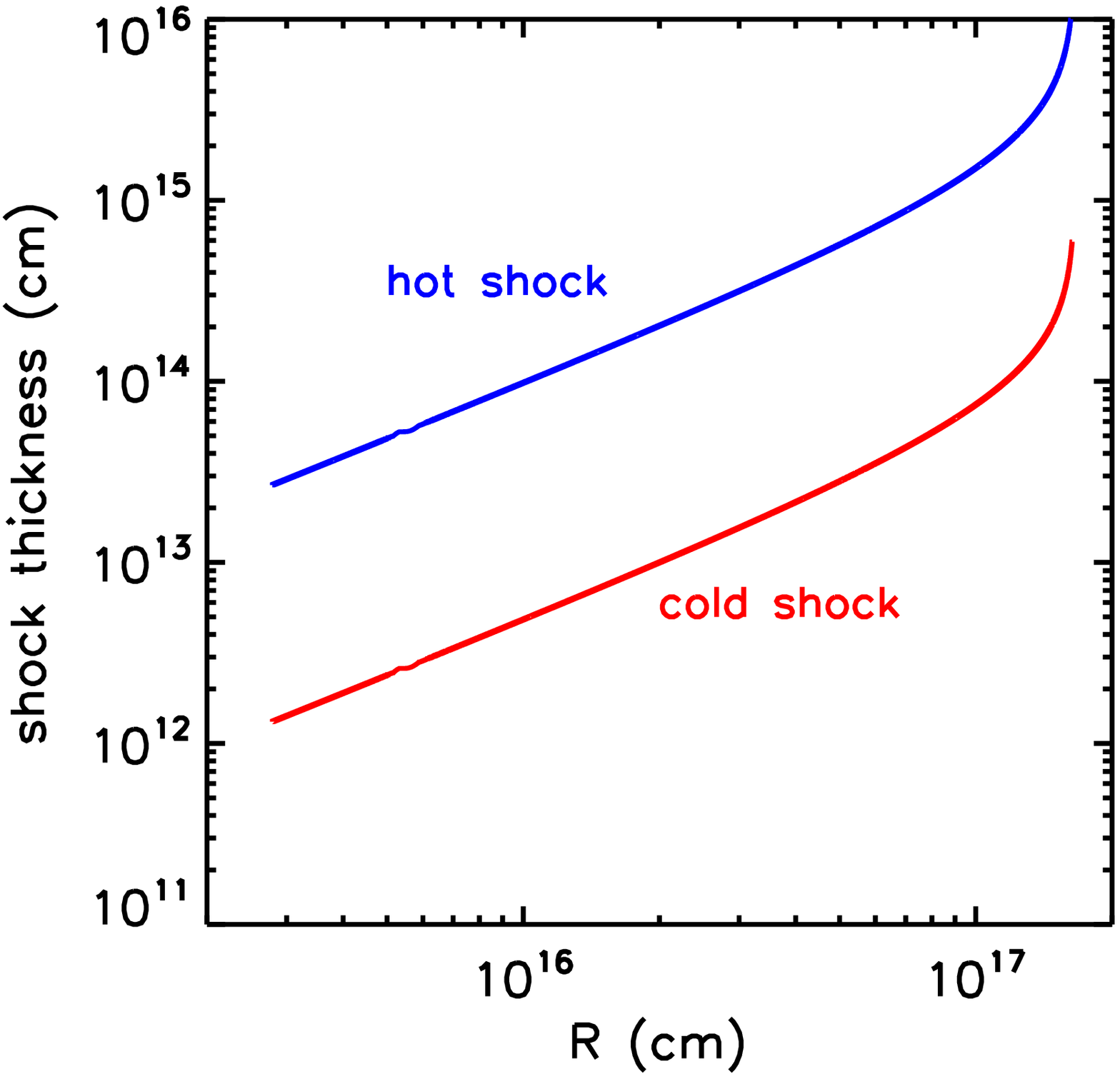}
\caption{The thickness of the cold and hot bow shocks are shown as a function of distance from \sgra. [The shock thickness of the cold bow shock 
scales approximately as $V_W^{-1.5} \dot{M}^{0.5}$ and 
as $V_W^{1/2} \dot{M}^{1/2}$ for the hot shock.]}\label{thickness}
\end{figure}

\begin{figure}[ht]
\epsscale{1.}
\plotone{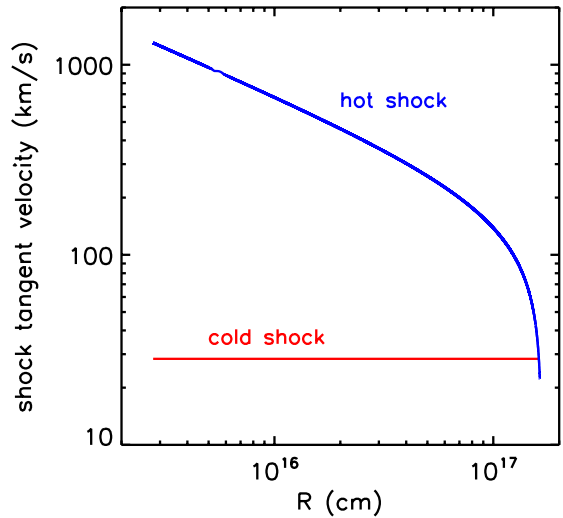}
\caption{The tangential velocities in the cold and hot bow shocks are shown as a function of distance from \sgra.  [The 
cold shock tangent velocity will scale linearly with $V_W$ and is independent of $\dot{M}$. The velocity in the hot shock is independent of both parameters.]}\label{tangental}
\end{figure}

The structure and physical conditions in the shocked envelope were computed over the full stellar orbit assuming 
stellar wind parameters of $V_{wind} = 100$ \kms, ${\dot M} = 4\times10^{-8}$ \msun~  yr$^{-1}$. Figures \ref{density} and \ref{thickness}  show the derived 
densities and thicknesses (perpendicular to the bow shock) for the two shocks as a function of the distance from \sgra. Over most of the orbit the 
cold shock density is 4 orders of magnitude higher than that of the hot shock, while the thickness of the hot shock is only $\sim20$ times that 
of the cold shock. It is therefore clear that the cold shock will have a much greater emission measure ($EM \propto n^2 vol$), provided the
gas is ionized. Figure \ref{tangental} shows the tangential flow velocities in the cold and hot show regions. (These velocities are parallel to the bow shock.)

\begin{figure}[ht]
\epsscale{1.}
\plotone{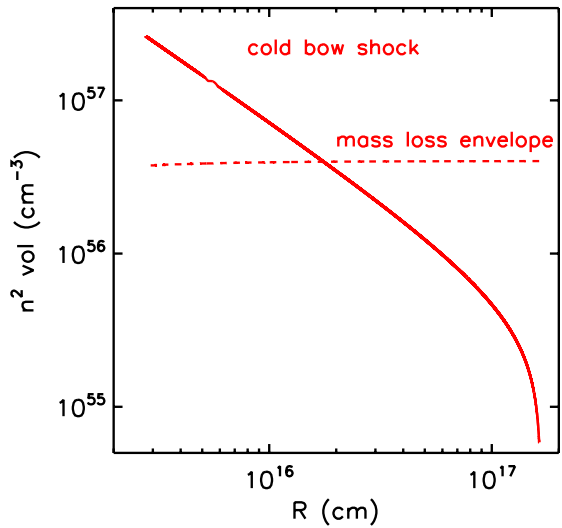}
\caption{The integrated  emission measure ($n^2 vol$) is shown for the cold bow shock and the stellar envelope as a function of distance from Sgr A$^*$. 
For the mass-loss envelope $n^2 vol$ is constant with a value determined by the adopted mass-loss, and very importantly, by
the inner radius adopted for the outflow (see discussion following Eq. \ref{rho_w}). 
At distance $10^{16}$ cm from Sgr A$^*$ in mid 2012, the total emission measure for the ionized gas is  $\sim10^{57}$ cm$^{-3}$.
[The emission measure from the cold shock 
scales as $V_W^{-1/2} \dot{M}^{3/2}$.] }\label{em}
\end{figure}

The gas in the hot shock layer flows around the cold shock layer tangentially at velocities $\sim1000$~\kms. The mass-flux of hot shock material intercepting the cold shock 
provides an upper limit to the cold shock ablation rate. For hot gas densities $\sim10^3$cm$^{-3}$ and an effective radius of 10$^{14}$ cm for the cold shock, this yields a maximum ablation rate of 10$^{-10}$\msun yr$^{-1}$
which is two orders of magnitude less than the wind mass-loss rate. Ablation is therefore probably not significant.

The derived emission measures (EM) for the cold shock and the mass-loss 
envelope integrated down to the launch radius ($\sim0.18$ AU for $V_{wind} = 100$\kms) are shown in Fig. \ref{em} as a function of distance from \sgra. For the bow shock, the EM 
was calculated only out to radius $\sim4\times10^{15}$, i.e. the area shown in Fig. \ref{model}. This clearly is a lower limit since there will be significant additional EM 
in the extended downstream tail of the bow shock. The EM 
values at $\sim10^{16}$cm (or 0.1\arcsec) from \sgra are in reasonable agreement with the Gillessen \etal (2012) value of $\sim10^{57}$ cm$^{-3}$ for the ionized gas, considering that we 
have not included the downstream tail. The EM of the bow shock is also $\sim4$ times that from the mass-loss envelope. 
The predicted EM of the bow shock will obviously increase if the mass-loss rate is raised above the adopted $4\times10^{-8}$ \msun~ yr$^{-1}$. 
In the foregoing discussion we have ignored the issue of whether the gas in the 3 zones will actually be ionized and clearly that is a critical consideration. 

\section{Ionization}\label{ionization}

Ionization of the material may potentially be provided by: UV photons from hot stars in the central few parsecs, UV/X-ray photons from 
the hot plasma in the central parsec or the hot gas ($10^{5-6}$ and $10^{8-9}$ K) in the two bow shocks, or collisions at  the inner bow shock where 
the envelope material moving at 100 - 500 km s$^{-1}$ is shocked. In the case of photoionization by extended UV sources such as the 
hot stars and the X-ray emitting plasma, it is important to recognize that it is not simply a matter of counting photons; one must also 
estimate the flux into the emission region. If the region of high EM is compact, only a small fraction of the available photons will actually be intercepted. 

\begin{figure}[ht]
\epsscale{1.}
\plotone{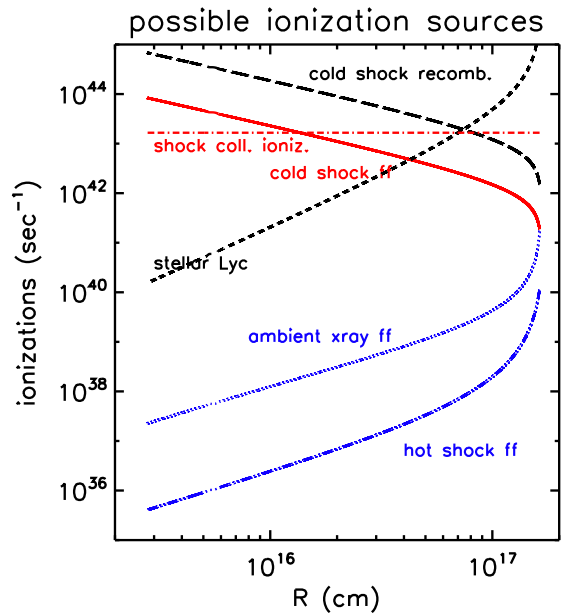}
\caption{The black dashed line shows the maximum hydrogen recombination rate (to $n > 1$) for the cold bow shock ($= n_e^2 vol \alpha_B$, i.e. assuming complete ionization of the shell) as 
a function of distance from \sgra, together with curves showing the possible sources of ionization:  collisional ionization of stellar wind material at the bow shock; free-free photons 
at energy greater than 13.6 ev from the hot X-ray emitting ambient medium, and from the hot and cold bow shock layers; and Lyman continuum photons from hot stars in the inner parsec.}\label{ion}
\end{figure}

In Fig. \ref{ion} we show the n$ > 1 $ (case B)  HI recombination rate for the cold bow shock material together with estimates for the possible ionization rates 
within the bow shock due to stellar Lyman continuum, free-free emission at greater than 13.6 eV from the hot plasma regions (the two bow shocks and the X-ray emitting 
ambient medium around \sgra) and collisional ionization due to mass-loss material passing through the bow shock . In the case of cooling radiation from the hot plasma regions
we have included only the free-free continuum, not the line cooling which dominates by a factor of a few at 10$^{5-6}$K, and for which some of the photons 
can ionize H \citep{shu79}. Although these ionization estimates are quite approximate, it is clear that the only viable sources are the collisional ionization at the bow shock 
and the gas cooling within the 'cold' bow stock where $T \sim10^{5-6}$ K. 

The Lyman continuum from the young stars in the galactic nucleus is not well constrained but 
we adopted the equivalent of a 
O5 star within the central 1 pc radius, i.e. a Lyman continuum production rate of Q = 10$^{50}$ Lyc s$^{-1}$. This production rate implies an average photon flux of $\sim3\times10^{12}$
Lyc cm$^{-2}$ s$^{-1}$ and integrating over the bow shock area we arrive at the estimate given in Fig. \ref{ion}. Although the production rate  
of  Lyman continuum photons from hot stars is large, the area of the bow shock is small, so very few of them will be intercepted. Similarly, although there is a potentially 
large EM at the base of the mass-loss envelope, the intercepted area of that region at radius 0.18 AU is so small that the material would not be ionized, except by UV from the stellar accretion shock. 
The latter may of course be substantial since the TTauri star winds exhibit ionized gas emission lines and some of the gas arriving at the bow shock may already be ionized.

\subsection{Constancy of the Emission Line Fluxes}

\cite{gil13} report that the observed line fluxes for the Br$\gamma$ emission are constant to within 10\% over the 4 year period
2008.5 to 2012.5. This is very surprising, given the fact that its distance from \sgra changed from 3.6 to 1.3$\times10^{16}$ cm and 
its 3d velocity increased from 1500 to 2900 \kms. In almost any model, one would expect the gas mass and/or the excitation of the emission in G2 to correlate with the distance from \sgra and/or 
infall velocity. In the context of the model proposed here, the approximate constancy of the flux might be understood if the dominant source of ionization 
is collisional, as stellar wind material from the inside passes through the cold bow shock at 100 to 500 \kms. In this case, the total number of ionizations per second (hence the line emission 
flux) will be constant and simply a few times the number of atoms passing through the shock front. This number flux is $\sim1.6\times10^{42}$ s$^{-1}$ for $\dot{M} = 4\times10^{-8}$ \msun~  yr$^{-1}$. 
Since the HI ionization energy of 13.6 eV corresponds to an HI particle velocity of 50\kms, this estimate should be multiplied by a factor $\sim(V_w /~50 \rm{\kms})^2$ to account for the energy available in the shock compared to what is needed to ionize HI. In Fig. \ref{ion}, we have 
taken this factor to be $\sim10$ for the collisional ionization rate estimate.
Combining this collisional ionization with the photo-ionization from free-free photons at $> 13.6$eV yields  an ionization rate ($\sim10^{44}$sec$^{-1}$), similar to what is needed to maintain the observed emission.  If the ionization is not due to collisions at the bow shock, it is unclear how to explain the constancy of the 
emission line fluxes since all the other sources of ionization vary with distance from \sgra. 

It is worthwhile noting that some variation in the emission line fluxes is a desirable feature of any model in which G2 comes in from larger radii. If the fluxes are 
relatively constant with radius, one should expect to see a large number of similar emission regions at the larger radii -- so far these have not been seen. A drop off in the expected 
emissivity out beyond several $10^{16}$ cm (as shown in Fig. \ref{em})  is therefore a desirable feature, in that it reduces the visibility of such precursors.

\section{The Star and its Orbit} 

\cite{eck13} report detection of an object they call DSO in the K and L band continua with position and proper motion similar to G2. The source has a very low color temperature, $\sim500$ K, and the 
K-band magnitude is $\sim18.9$ mag. For 1-2 \msun~ TTauri stars,  M$_K \sim 2 - 4$ mag \citep[][, Hillenbrand -- private communication 2013]{bar98}. At the Galactic center, the 
un-extincted apparent magnitude will be 16.5 - 18.5 mag in K-band. The K-band extinction towards the Galactic center will dim it a further $\sim3$ mag, implying $m_K \sim20$ mag. 
Thus,  the star itself, if it is a TTauri star, would be very difficult to directly detect unless it is in a period of enhanced activity. In fact, \cite{eck13} have suggested that G2/DSO
is a dust enshrouded star.  One would also expect there to be mass-loss red giant stars in the galactic nucleus; however, such stars can probably be ruled out for G2
since their brightness would be higher than the observed L-band flux. 

The origin of the high eccentricity orbit of G2 remains poorly understood. Although one might posit stellar interactions in a triplet system, these would generally be disruptive 
of any circumstellar material. The injection to a highly eccentric orbit  from circular orbit 
requires a velocity kick of $\sim500$ \kms. To provide such an impulse with a single star-star scattering would require a close approach well inside 1 AU. Such close 
encounters would certainly disrupt any large protoplanetary disk such as that invoked by \cite{mur11} -- and possibly also the inner disk from which the TTauri star winds are launched (as discussed here). 

In the face  
of such difficulties, it is attractive to consider the possibility that the young stars would have to be formed in eccentric orbits through collisions of gas clumps with cancelation of angular momentum (Alig \etal 2013 -- in preparation). The gas motions in the mini-spiral inside 1 parsec have substantial non-circular velocities and the molecular ring or CND at 1-3 parsec 
radius is inclined at 50 - 75\deg to the Galactic plane  \citep{jac93,chr05} -- both of which indicate substantial non-planar, non-circular dynamics for the ISM there. 

Alternatively, the increase in eccentricity might be built up 
by a series of many smaller amplitude scatterings \citep[e.g.][]{mur11}.
Is it possible that a large increase in the eccentricity from an initial circular orbit could be induced similar to the Kozai 
oscillations in exo-planetary systems? If the star was formed in the young stellar ring at $2\times10^{17}$ cm radius, the orbital 
period is $\sim200$ yr. The star and any companions could have then orbited the galactic center $\sim10^4$ times. Mass clumps associated with both the circumnuclear gas disk and stars might possibly provide perturbations to initiate the process. 

\section{Implications} 

\begin{figure}[ht]
\epsscale{1.}
\plotone{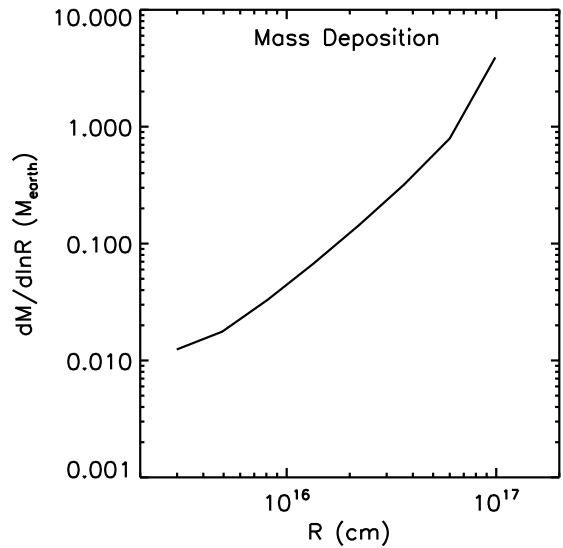}
\caption{The mass deposition from the stellar mass loss is shown as a function of distance from \sgra, obtained by integrating the mass-loss along the orbit.}\label{mass}
\end{figure}

The usually assumed mass for G2 of 3 M$_\oplus$ was derived by \cite{gil12a} assuming the gas is distributed homogeneously 
with density $\sim6\times10^5$cm$^{-3}$. For our model the emission measure is produced by gas at density $\sim10^8$cm$^{-3}$, 
resulting in a decrease in the required mass of emitting gas by a factor $\sim100$. Fig. \ref{mass} shows the mass deposition from 
the stellar mass loss as a function of radius from \sgra. Approximately 0.1 M$_\oplus$ is deposited in the vicinity of \sgra.

It would be a shame if this object, which so intrigues us now, were to disappear this September at pericenter. The model proposed 
here looks to a brighter future. At pericenter, the tidal radius is reduced to $\sim1 - 3$ AU -- this major disruption in the mid-radii of the 
disk will result in stripping to the exterior and some deposition to the interior disk -- inside 1 AU. The latter 
could result in greatly enhanced stellar mass-loss rates -- hence much brighter emission at and after pericenter passage for several 
years.

\acknowledgments

We would like to thank Lynne Hillenbrand for  helpful discussions on the properties of TTauri stars and Andreas Eckart for suggestions. AB thanks Caltech for the hospitality of a visit which stimulated this project.  We acknowledge discussions with Alessandro Ballone and Marc Schartmann who are 
further developing this model with detailed numerical simulation.

\bibliography{scoville_g2}

\end{document}